\documentstyle[prl,aps,epsf]{revtex}
\newcommand{\bv}[1]{{\bf #1}}
\begin{document}

\title{Optical detection of a BCS phase transition in a trapped 
gas of fermionic atoms}
\author{Weiping Zhang$^{1,2}$, C. A. Sackett$^2$, and R. G. Hulet$^2$} 
\address{$^1$Department of Physics, Macquarie University, Sydney, NSW 2109,
Australia}
\address{$^2$Physics Department and Rice Quantum Institute,
Rice University, Houston, Texas 77251}
\date{October 1, 1998}
\maketitle
\begin{abstract}
Light scattering from a spin-polarized degenerate Fermi gas of trapped 
ultracold $^6$Li atoms is studied.
We find that the scattered light contains information which directly reflects the
quantum pair correlation due to the formation of atomic Cooper pairs resulting
from a BCS phase transition to a superfluid state.  Evidence for pairing
can be observed in both the space and time domains.
\end{abstract}
\pacs{03.75.Fi, 32.80.-t, 32.80.Lg}
\widetext

The realization of Bose-Einstein condensation in trapped 
atomic gases \cite{BEC} has
generated interest in the atomic physics, quantum optics and 
condensed matter physics communities. Although the experimental realization
of  a degenerate atomic Fermi gas has
not yet been demonstrated, interest in this subject is  
increasing \cite{Stoof96,Bruun98,Baranov96,Rokhsar97}. Of course,
the behavior of a degenerate Fermi gas is remarkably different from
a degenerate Bose gas. By analogy with the BCS theory of superconductivity
in metals, it has been predicted that a degenerate  
Fermi gas can undergo a BCS phase transition to an atomic
superfluid state if the interatomic interaction in the gas
is attractive \cite{leggett}.  
Experiments to trap and cool $^6$Li and $^{40}$K gases 
into the quantum degenerate regime 
are underway in several laboratories.

In this paper, we address the question of how to detect the superfluid state 
after the BCS phase transition. 
We assume that the Fermi gas has been cooled to near
absolute zero, so that all trap levels up to the Fermi energy are filled. 
An attractive interatomic interaction will cause atoms in the
vicinity of the Fermi level to form Cooper pairs, with each pair composed of 
two quantum correlated atoms behaving as a new composite Bose
particle. These bosons automatically undergo Bose-Einstein condensation and
form a superfluid. The quantum pair correlation of the
Cooper pairs characterizes the superfluid properties of the gas. 

A promising experimental approach is to prepare a degenerate gas
with atoms in an incoherent mixture 
of two internal hyperfine state.  Such a mixture
allows Cooper pairing via an {\it s}-wave interaction, and
leads to practically attainable BCS-transition temperatures when 
the scattering length {\it a}
is large and negative. This occurs naturally for $^6$Li \cite{slength}, or 
can be obtained in the
vicinity of a Feshbach resonance for other atoms\cite{Verhaar}. 
We consider here a trapped $^6$Li gas 
in an incoherent mixture of ground states 
$|+ \rangle = |M_s=1/2,M_I=1 \rangle$ and $|- \rangle = |M_s=1/2,M_I=0
\rangle$ \cite{Stoof96}. 

The key to observing the superfluid state is to determine the existence of pair
correlations. To achieve this goal, we propose to use
off-resonance light scattering and Fourier imaging techniques. 
A laser beam with amplitude $\bv{E}_L$, 
frequency $\omega_L$, and wave vector $k$ propagating
along the $z$ direction is used to illuminate the gas. 
We take the light to be linearly polarized and tuned near resonance
between an $S$ ground state and $P$ excited state.  
To avoid incoherent heating of the gas
due to spontaneous emission, the magnitude of the laser detuning,
$\delta=\omega_L-\omega_0$, is assumed to be large.
In vector quantum field theory \cite{vecqf1,vecqf2,vecqf3}, the atoms
in the light field can be described by a four-component 
atomic field $\Psi(\bv{r})=\psi_{+} |+\rangle +\psi_{-} |-\rangle + 
\psi_{e+} |e+\rangle +\psi_{e-} |e-\rangle$ with $\psi_{\pm}$ denoting
atoms in the ground-state hyperfine levels $|\pm \rangle$, and 
$\psi_{e \pm}$ in the corresponding
excited-state hyperfine levels. For large $\delta$, the excited-state 
components can be adiabatically eliminated, yielding a total atomic 
polarization operator with positive-frequency part \cite{vecqf1} 
\begin{eqnarray}
\bv{P}^{(+)}(\bv{r},t) = -\wp \frac{\wp \cdot \bv{E}^{(+)}}
{
	  \hbar \delta} \hat{\rho}(\bv{r},t) e^{-i \omega_L t},
\label{polarization} 
\end{eqnarray}
where $\hat{\rho}(\bv{r},t)=\psi^{\dagger}_{+}(\bv{r},t)
\psi_{+}(\bv{r},t) 
+ \psi^{\dagger}_{-}(\bv{r},t) \psi_{-}(\bv{r},t)$ denotes the total
atomic density
operator in the ground state, $\wp$ the matrix element of the atomic
dipole moment, and $\bv{r}$ a location in the gas. Light propagation 
is determined by the atomic polarization operator~(\ref{polarization}) and
the wave equation
\begin{eqnarray}
{\nabla}^2 \bv{E}^{(+)}-\frac{1}{c^2} \frac{{\partial}^2 \bv{E}^{(+)}} 
{\partial t^2} &=& {\mu}_0 \frac{{\partial}^2 \bv{P}^{(+)}} {\partial t^2}.
\label{maxwell} 
\end{eqnarray}
The solution to Eq.~(\ref{maxwell}) can be expressed as
\begin{eqnarray}
\bv{E}^{(+)}(\bv{R},t) &=& \bv{E}_S^{(+)}(\bv{R},t) e^{-i \omega_L
t}
+ \bv{E}_L^{(+)} e^{ik z-i \omega_L t}, 
\label{solution} 
\end{eqnarray}
where $\bv{E}_S^{(+)}(\bv{R},t)$ is the scattered field at position 
$\bv{R}$.
For $R \equiv |\bv{R}| \gg |\bv{r}|$,
the scattered field has the form
\cite{vecqf1,vecqf3} 
\begin{eqnarray}
\bv{E}_S ^{(+)}(\bv{R},t) = k^2 \frac{e^{i k R}} {R} \int d^3 r 
e^{-i k \hat\bv{R} \cdot \bv{r}} 
\Big[\bv{P}^{(+)}(\bv{r},t)-\hat\bv{R} \cdot
\bv{P}^{(+)}(\bv{r},t) \hat\bv{R} \Big],
\label{scatter} 
\end{eqnarray}
where the directional unit vector $\hat\bv{R} = \bv{R}/R$. 
From Eqs.~(\ref{polarization})~and~(\ref{scatter}), we see that
the scattered field depends on the density operator of the gas, so that
the averaged spectral intensity of the scattered field received by a 
photodetector contains the second-order correlation of 
the atomic field operators\cite{photodetector}
\begin{eqnarray}
\langle \hat{\rho}(\bv{r},t) \hat{\rho}(\bv{r}^{\prime},t^{\prime}) \rangle
\approx 
\langle \hat{\rho}(\bv{r},t) \rangle  \langle
\hat{\rho}(\bv{r}^{\prime},t^{\prime}) 
\rangle + G(\bv{r},\bv{r}^{\prime},t,t^{\prime}),
\label{correlation} 
\end{eqnarray}
where ``$\langle \ldots \rangle$'' 
denotes the quantum mechanical expectation value.
The first term in Eq.~(\ref{correlation}), which depends on the total 
averaged density, describes the contribution to the scattered field
by the normal ground-state component. The second term,
\begin{eqnarray}    
G(\bv{r},{\bv{r}}^\prime,t,t^{\prime}) \equiv 
-2 \langle \psi_{-}(\bv{r},t) \psi_{+}(\bv{r}^{\prime},t^{\prime}) \rangle 
\langle \psi_{-}^{\dagger}(\bv{r},t)
\psi_{+}^{\dagger}(\bv{r}^{\prime},t^{\prime}) 
\rangle,
\label{correlation1} 
\end{eqnarray}
gives the quantum pair correlation function arising from the formation of
Cooper pairs in the superfluid state.

The contribution of the laser field $\bv{E}_L$ in Eq.~(\ref{solution})
can be removed by imaging the cloud with a dark ground technique,
as discussed in Refs.~\cite{darkground}.  
If a plane located a distance $z_0$ from 
the atoms is observed in this way, the spectral and spatial intensity
distribution measured on the detector will be \cite{photodetector} 
\begin{eqnarray}
I(\bv{R}_\perp,\nu) =\int_{-\infty} ^{\infty} d\tau 
e^{i \nu \tau} \frac{1}{2T} \int_{-T} ^T dt 
\langle \bv{E}_{S}^{(-)}(\bv{R}_0,t) \cdot 
\bv{E}_{S}^{(+)}(\bv{R}_0,t+\tau) \rangle,
\label{specint} 
\end{eqnarray}
where 2T is the time interval used for detection, and 
$\bv{R}_0 \equiv (\bv{R}_{\perp}, z_0)$ is a point in the image plane.
Equation~(\ref{scatter}), along with relations
(\ref{polarization}) and (\ref{correlation}), gives the spatial-temporal
correlation function of the light field
\begin{equation}
\langle \bv{E}_{S}^{(-)}(\bv{R}_{0},t) \cdot 
\bv{E}_{S}^{(+)}(\bv{R}_{0},t+\tau) \rangle       
= \frac{9 I_L {\gamma}^2}{16 (k z_0 \delta)^2}
\Big[I_1(\bv{R}_\perp,t,\tau) + I_2(\bv{R}_\perp,t,\tau) \Big] 
e^{-i \omega_L \tau},
\label{Ep2} 
\end{equation}
where $I_L = \bv{E}_L^{(-)} \cdot \bv{E}_L^{(+)}$ is the intensity of the
incident light and $\gamma$ is the natural linewidth of the transition.
The functions $I_1$ and $I_2$ are defined as 
\begin{eqnarray}
I_1(\bv{R}_{\perp},t,\tau) = \int\int 
d^2 \bv{r}_{\perp} d^2 \bv{r}_{\perp}^{\prime} 
e^{-i k \bv{R}_\perp \cdot (\bv{r}_\perp
	- \bv{r}_\perp^{\prime})/z_0 }
\langle \hat{\rho}(\bv{r}_\perp,t) \rangle 
\langle \hat{\rho}(\bv{r}_\perp^{\prime},t+\tau) \rangle, 
\label{I1}
\end{eqnarray}
and
\begin{eqnarray}
I_2(\bv{R}_{\perp},t,\tau) = \int d^2 \bv{\xi} 
e^{-i k \bv{R}_\perp \cdot \bv{\xi}/z_0 }
\int d^2 \bv{r}_\perp G(\bv{r}_\perp,\bv{r}_\perp -
\bv{\xi},t,t+\tau),
\label{I2} 
\end{eqnarray}
where the relative distance between atoms is denoted by
$\bv{\xi}=\bv{r}_{\perp}-\bv{r}_{\perp}^{\prime}$. The function $I_1$
describes the signal from the normal component of the gas
and $I_2$ the signal from the Cooper pairs.  In general,
$I_2$ is much weaker than $I_1$ since the averaged density of atoms in the 
normal component is far larger than that of the pairs. 

The averaged density and the quantum pair correlation
function can be found using vector quantum field theory \cite{vecqf1}.
In the off-resonant light field, the degenerate Fermi
gas is described by the coupled quantum field equations
\begin{eqnarray} 
i \hbar \frac{\partial \psi_{+}}{\partial t}
=(H_0 -\mu_{+} + V_L-i \hbar \Gamma/2) \psi_{+} 
- \Delta(\bv{r}) \psi_{-}^{\dagger}
\nonumber \\ 
i \hbar \frac{\partial \psi_{-}^{\dagger}}{\partial t}
=-(H_0 -\mu_{-} + V_L + i \hbar \Gamma/2) \psi_{-}^{\dagger} 
- \Delta(\bv{r}) \psi_{+},
\label{EH}
\end{eqnarray}  
where $H_0=-\frac{\hbar^2}{2m}\nabla^2 + \frac{1}{2}m \omega^2 r^2$ is
the free Hamiltonian of the trapped Fermi gas, $V_L = 
\hbar \Omega^2 /4 \delta$ is the light-induced 
potential, $\Gamma = \gamma \Omega^2 /4 \delta^2$ 
is the rate for spontaneous emission,
$\mu_{\pm}$ are the chemical potentials of the two internal states,
and $\Delta(\bv{r}) = (4 \pi |a| \hbar^2/m) \langle 
\psi_{-}(\bv{r}) \psi_{+}(\bv{r}) \rangle$
is the BCS energy gap function\cite{Stoof96}. The Rabi frequency of the
light field is $\Omega \equiv |\wp \cdot \bv{E}_L /\hbar|$.
For simplicity we consider the simple case $\mu_{+}=\mu_{-}$ 
for equal number of atoms in each spin state and introduce the
renormalized chemical potential $\mu=\mu_{+}-V_L$. Further, a large laser 
detuning and a weak intensity allow 
$\Gamma \ll \mu, \Delta$ so that destruction of Cooper pairs by
spontaneous emission and interactions involving excited-state 
atoms \cite{vecqf1}
can be neglected. Employing an approach similar to that 
adopted in BCS theory, we approximate the solutions of Eqs.~(\ref{EH})
by
\begin{eqnarray} 
\psi_{ \pm}(\bv{r},t)= \sum_{\bv{n}} \Big(u_{\bv{n}}(\bv{r})
\hat{b}_{\bv{n} \pm} e^{-iE_{\bv{n}}t/\hbar} \pm v_{\bv{n}}(\bv{r})   
\hat{b}_{\bv{n} \mp}^{\dagger} e^{iE_{\bv{n}}t/\hbar} \Big),
\label{BT}
\end{eqnarray} 
where $\hat{b}_{\bv{n} \pm}$ are generalized Bogoliubov quasi-particle
operators and $E_\bv{n}$ the excitation energy
for the mode indexed by $\bv{n}$. 
The superfluid state of the degenerate Fermi gas 
is characterized by the BCS ground state $|\Phi_{BCS} \rangle$ with 
the property $\hat{b}_{\bv{n} \pm} |\Phi_{BCS} \rangle=0$. 
From Eqs.~(\ref{EH}) with the dissipative terms ignored, the transformation 
coefficients $\{ u_{\bv{n}},v_{\bv{n}} \}$ satisfy the celebrated
Bogoliubov equations
\begin{eqnarray} 
(H_0 -\mu) u_{\bv{n}}(\bv{r}) + \Delta(\bv{r}) v_{\bv{n}}(\bv{r})
=E_{\bv{n}} u_{\bv{n}}(\bv{r}) 
\nonumber \\ 
-(H_0 -\mu) v_{\bv{n}}(\bv{r}) + \Delta(\bv{r}) u_{\bv{n}}(\bv{r})
=E_{\bv{n}} v_{\bv{n}}(\bv{r}).
\label{BE}
\end{eqnarray}  
The total averaged density can be expressed as 
$\langle \hat{\rho}(\bv{r},t) \rangle \equiv \langle \Phi_{BCS}|
\psi_+^{\dagger} \psi_+^{\mbox{}} +\psi_-^\dagger \psi_-^{\mbox{}}
|\Phi_{BCS} \rangle = 2 \sum_{\bv{n}} |v_{\bv{n}}(\bv{r})|^2$ and 
the quantum pair function is
\begin{eqnarray}
G(\bv{r},\bv{r}^{\prime},t,t^{\prime})=2
\sum_{\bv{n} \bv{m}} u_{\bv{n}}(\bv{r})
v_{\bv{n}}(\bv{r}^{\prime})
u_{\bv{m}}(\bv{r}^{\prime}) v_{\bv{m}}(\bv{r}) 
e^{-i(E_{\bv{n}}+E_{\bv{m}})(t-t^{\prime})/\hbar}.
\label{QPC}
\end{eqnarray}  

The average density and pair function can be 
calculated by self-consistently solving Eqs.~(\ref{BE}).  
In the normal degenerate ground state, energy levels below the
Fermi level $E_F$ are occupied, while those above are empty.
The effect of interatomic interactions is to cause scattering 
between nearby energy levels, which
creates an energy shell near $E_F$ where normally
unoccupied states in the normal ground state acquire an amplitude 
to be occupied, and states below $E_F$ have some amplitude
to be unoccupied.  The stronger the interatomic interaction is, the wider
the energy shell and the more atoms are available to form Cooper pairs. 
Physically, the coefficients 
$u_{\bv{n}}$ and $v_{\bv{n}}$ in Eqs.~(\ref{BE}) determine the
amplitudes for atoms to be scattered into the pair states.  To evaluate these
amplitudes, we expand the coefficients as $u_{\bv{n}}=\sum_{\bv{q}} 
u_{\bv{n} \bv{q}} \phi_{\bv{q}}$ and $v_{\bv{n}}=\sum_{\bv{q}}
v_{\bv{n} \bv{q}} \phi_{\bv{q}}$, in terms of  
the eigenstates $\phi_{\bv{q}}$ of the single-atom Hamiltonian $H_0$.
In principle the sum over $\vec{q}$ in the coefficients should extend  
from zero to infinity. However, we should note that in the BCS theory
\cite{Stoof96,Bruun98}, Equation (13) is a direct result of the Born approximation 
by replacing the realistic non-local interatomic interaction 
$V(\vec{r})$ by a local contact potential $V(\vec{r})=4 \pi \hbar^2 
a \delta(\vec{r})/m$. 
We know that the Born approximation is only valid for low-energy scattering. 
The invalidity of the approximation in high-energy scattering regime 
produces an ultra-violet divergence in the BCS theory. In the case of
superconductivity, the ultra-violet divergence naturally vanishes by 
considering the fact that the phonon-exchange induced interaction between
electrons can be cut-off in the Debye frequency. However, in the 
case of degenerate Fermi gas of atoms, to avoid the ultra-violet 
divergence, an exact theory for superfluid phase transition
must take the realistic
shape of the exact non-local triplet potential into account. 
Recently two independent approaches to remove the ultra-violet divergence 
in the BCS theory of degenerate Fermi gas of atoms have been proposed 
\cite{Stoof96,Bruun98}. One is to renormalize the interaction potential 
in term of the 
Lippman-Schwinger equation \cite{Stoof96} and the other is to employ the more 
exact pseudo-potential approximation \cite{Bruun98}.
However for a first guess, the Born approximation provides a simple and 
reasonable way to evaluate the gap energy and the pair correlation 
if an appropriate momentum cut-off is introduced to remove the 
ultra-violet divergence. Now the question is how to choose a physically 
valid momentum cut-off $\hbar k_c$. To determine the cut-off range, we 
must use the fact that Born approximation only gives the correct evaluation 
in the low-energy scattering regime with $k|a| < 1$. Hence the validity
of the present theory based on Born approximation requires a cut-off 
$k_c < |a|^{(-1)}$. For $^6Li$ atom, this is in the order of Fermi wave 
number $k_F$. With such a cut-off, we numerically evaluate the energy gap,
the total averaged atomic density and the quantum pair function. 

\begin{figure}
\epsfxsize=11cm
\epsffile{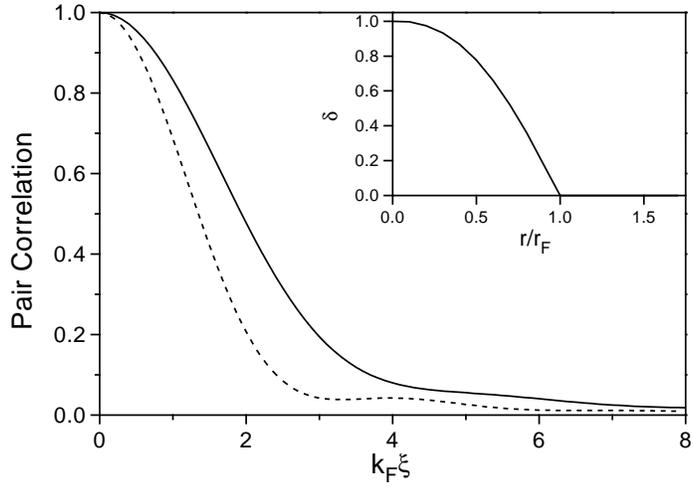}
\caption{
The normalized spatial distributions of the equal-time quantum pair 
correlation function, with the distance $\xi$ between atoms scaled 
by the Fermi wavenumber $k_F$. The dashed curve is the 
correlation due to Cooper pairs with center of mass located at the 
trap center and the solid curve is the average contribution of all Cooper pairs.
The inset shows the spatial dependence of the energy gap, normalized to
the center of the trap and with position scaled by the Fermi radius $r_F$. 
}
\label{pair}
\end{figure}

To be concrete, we assume that $N=2 \times 10^5$ $^6$Li atoms in each spin
state are confined by a magnetic trap with an oscillation frequency
$\omega=2 \pi \times 150$ Hz.  With these values, 
$E_F \approx 100\hbar\omega \approx 740$~nK,
and the peak value of the energy gap is 
$\Delta(0) \approx 5 \hbar \omega =36$~nK\@.  For a 
degenerate Fermi gas in a harmonic trap, 
the characteristic size of the average density is given by
the Fermi radius $r_F = [2 E_F/m \omega^2]^{1/2} \approx 48~\mu$m \cite{size},
while the length scale of the
pair correlation function is $r_c \sim {k_F}^{-1}$, where
$k_F =(2mE_F/\hbar^2)^{1/2} \approx 2\pi \times 6800$~cm$^{-1}$ 
is the Fermi wavenumber. The numerical result for the correlation function 
is shown in Fig.~\ref{pair}, 
along with the spatial variation of the energy gap.

We need emphasize that in the homogeneous gas,
the correlation length (pair size) at zero temperature
is defined in terms of the so-called
coherence length $\xi_c =\hbar v_F /\Delta(0)$.  
The coherence length determines the region where the pair function
extends \cite{leggett}. However within the region, 
the pair function still contains
shorter oscillation structure which has the scale $r_c$. In fact
from our numerical result for the trapped gas, we see that the  
pair function indeed varys with such a length scale.
Now we will explain how such a scale can be observed by
optical imaging.

\begin{figure}
\epsfxsize=13cm
\epsffile{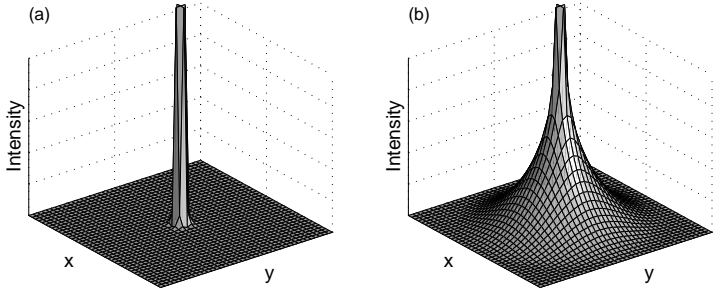}
\caption{
The spatial image measured a distance $z_0$ from the atoms, with
(a) the trapped Fermi gas in the normal degenerate ground state
and (b) in the superfluid state after the BCS phase transition.
In both images, the central peak is clipped and actually extends by a
factor of $\sim$$10^6$ above the axes shown.  The radial size of the
normal component is approximately $z_0/kr_F$, while the pair component
is larger, extending to $z_0/kr_c$.
}
\label{image}
\end{figure}

\begin{figure}
\epsfxsize=11cm
\epsffile{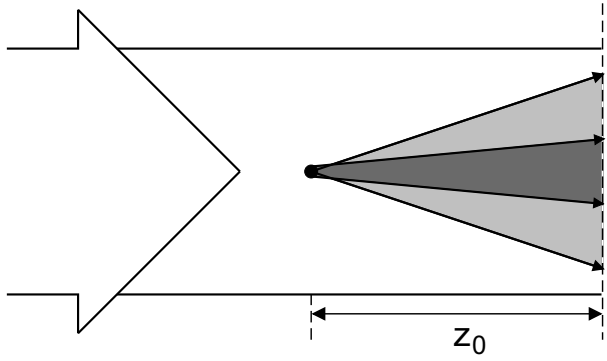}
\caption{\label{schematic}
Schematic of the imaging technique.  The white area represents the
incident probe laser, the dark gray the light coherently scattered
by the cloud, and the light gray the light scattered by Cooper pairs.
The small length scale of the pair structure scatters light at a relatively
large angle, so by measuring the intensity in the far field, the 
components can be distinguished.  Dark ground imaging techniques can be used
to eliminate the contribution of the probe laser 
itself \protect\cite{darkground}.}
\end{figure}

Assuming that a plane at $z_0$ = 2 cm is imaged with
unit magnification and with a transition wavelength $\lambda =670$ nm,
the image size is $z_0 /k r_F \sim 0.09$ mm for the normal component
and $z_0 /k r_c \sim 1.9$ cm for the pair component, differing by a factor
of $2E_F/\hbar\omega$.  The calculated
images for a gas below and above the critical temperature for the BCS 
phase transition are shown in Fig.~\ref{image}(a) and (b), respectively.
It is seen that when the transition occurs, a spatially broadened image 
appears.  The physical situation is depicted in Fig.~\ref{schematic}, where
the small-scale structure induced by pairing causes light to scatter
at a larger angle than that scattering from the cloud itself.

The normal signal is produced by coherent scattering and is therefore
proportional to $(2N)^2$, as can be verified by reference to Eq. (9).  The pair
signal, however, arises from spontaneous Raman scattering between 
pairs above and below the energy gap,
and is found using Eq. (10) to
be proportional to the number of pairs $N_p$.  This number is determined by the
number of atoms in an energy shell of width $\Delta$ centered on $E_F$, so
$N_p \approx 3 N \Delta / E_F$.  For the parameters given above, 
$N_p \approx 3 \times 10^4$ and the ratio of the peak signal
intensities is $I_2(0) / I_1(0) \approx 2 \times 10^{-7}$.  
It is difficult to experimentally measure a signal with such a large dynamic
range, but the pair signal can be revealed by using a nearly opaque 
spatial filter to attenuate the normal signal.  
If the diameter of the filter is chosen to be
approximately equal to the spatial dimension of the normal signal image, it 
will affect only the central region of the pair signal, and both contributions
can be observed with the same intensity scale.  

\begin{figure}
\epsfxsize=11cm
\epsffile{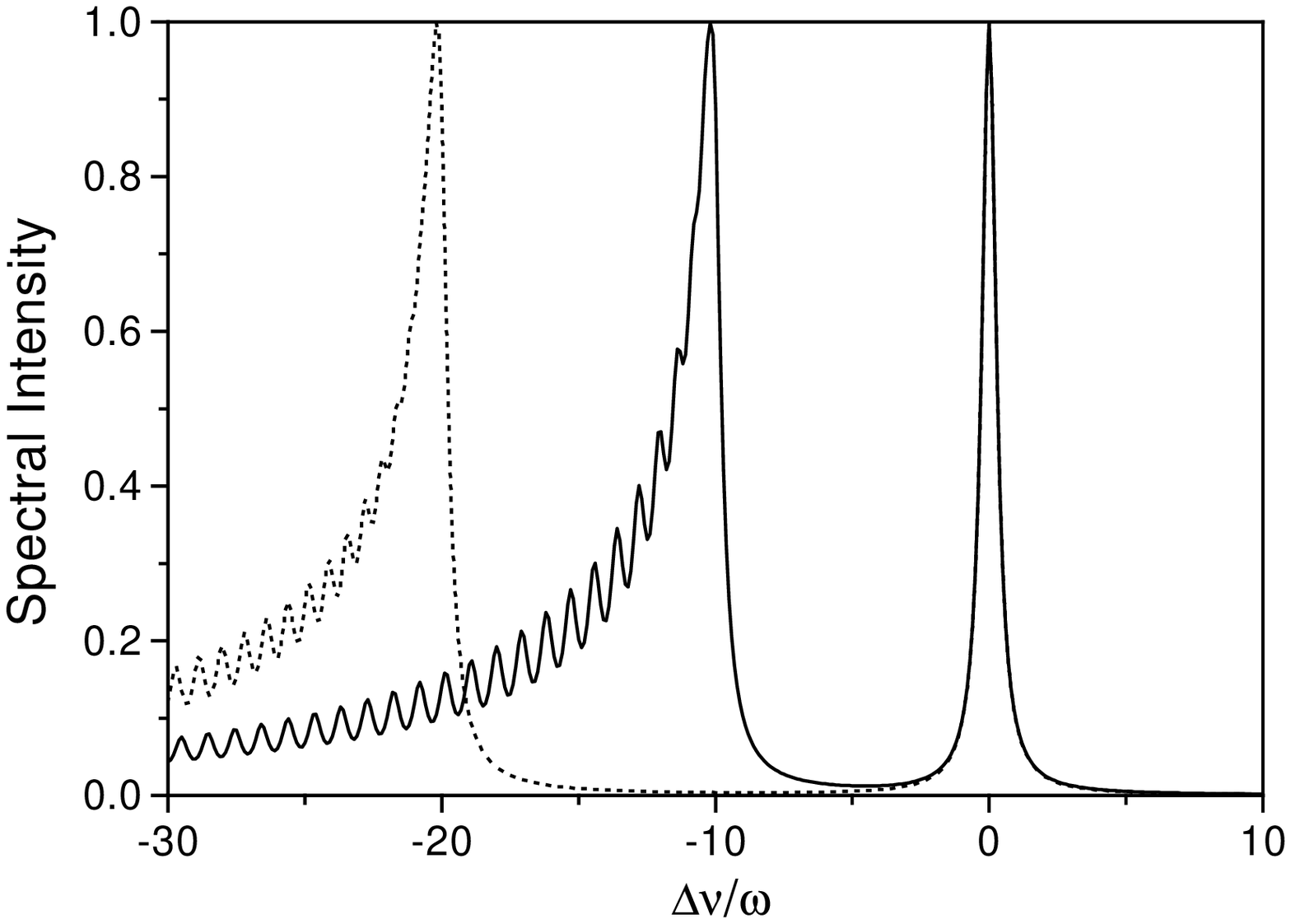}
\caption{
The normalized scattered light spectrum of the scattered field.
The frequency shift $\Delta \nu$ is scaled by the trap 
frequency $\omega$.  For the solid curve, the gap energy
$\Delta \approx 5~\hbar\omega = 36$~nK, and for the dotted curve 
$\Delta = 72$~nK.
A spatial filter with a transmission of $\sim$$10^{-4}$ is used
to reduce the strong signal from the normal component to the same level as 
that from the pair component.
}
\label{spectrum}
\end{figure}

Finally, we calculate the scattered light 
spectrum. For the normal degenerate ground state, a single
spectral line is obtained
at the frequency of the incident light. For the superfluid state, the
spectrum exhibits a double-peaked 
structure as shown in Fig.~\ref{spectrum}. The coherent peak
is from scattering by the normal component.  
The frequency shift of the sideband line is approximately twice the gap
energy, confirming that the sideband is due to Raman scattering by pairs. 
The long oscillating tail of the sideband is due to modulated broadening from
the center of mass motion of atoms at the trap frequency.
Hence, the presence of the shifted peak provides another effective 
method to detect the BCS phase transition and can be 
used to directly determine the gap energy.

The theory presented here was simplified by the neglect of spontaneous
emission, permitting, for example, the assumption that $\Delta$ remains constant
during probing.  However, the pair signal depends on breaking pairs by
incoherent spontaneous Raman scattering, and thus requires spontaneous
emission.  The theory is therefore valid only in the weak-signal limit,
where $\Gamma \ll T^{-1}$.  Larger signals could be obtained
experimentally by allowing $\Gamma \sim T^{-1}$, 
but quantitative interpretation would then be more difficult.

In conclusion, we have studied off-resonance light scattering by a trapped
degenerate Fermi gas.
The results show that both spatial imaging and the scattered light spectrum
give clear signatures for the BCS phase transition to a gaseous 
superfluid state.

The work in Australia was supported by the Australian Research Council,
and a Macquarie University Research Grant.  The work at Rice was supported
by the NSF, ONR, NASA, and the Welch Foundation.  WZ thanks the
atom-cooling group at Rice for their hospitality during his visit and also 
thanks Karl-Peter Marzlin for his help.


\newpage

\newpage

\newpage

\newpage

\begin{thebibliography}{99}

\bibitem{BEC} M. Anderson, J. R. Ensher, M. R. Matthews,
C.~E. Wieman, and E.~A. Cornell, Science {\bf 269}, 198 (1995);
C.~C. Bradley, C.~A. Sackett, J.~J. Tollett, and R.~G. Hulet,
Phys. Rev. Lett. {\bf 75}, 1687 (1995);
K.~B. Davis, M.-O. Mewes, M.~R. Andrews, N.~J. van Druten, D.~S. Durfee,
D.~M. Kurn, and W. Ketterle, Phys. Rev. Lett. {\bf 75}, 3969 (1995).

\bibitem{Stoof96} H.~T.~C. Stoof, M. Houbiers, C.~A. Sackett, and
R.~G. Hulet, Phys. Rev. Lett. {\bf 76}, 10 (1996);
M. Houbiers, R. Ferwerda, H.~T.~C. Stoof, W.~I. McAlexander, C.~A.
Sackett, and R.~G. Hulet, Phys. Rev. A {\bf 56}, 4864 (1997).

\bibitem{Bruun98} G. Bruun, Y. Castin, R. Dum, and K. Burnett,
preprint (cond-mat/9810013) 1998.
 
\bibitem{Baranov96} M.~A. Baranov, Yu. Kagan, and M.~Yu. Kagan, Pis'ma
Zh. Eksp. Teor. Fiz. {\bf 64}, 273 (1996) [JETP Lett. {\bf 64}, 301 (1996)].

\bibitem{Rokhsar97} D.~A. Butts and D.~S. Rokhsar, Phys. Rev. A {\bf 55}
(4346) 1997.

\bibitem{leggett} A.~J. Leggett, J. Phys. (France) IV {\bf C}7, 19 (1980).

\bibitem{slength} E.~R.~I. Abraham, W.~I. McAlexander, J.~M. Gerton,
R.~G. Hulet, R.~C\^{o}t\'{e}, and A. Dalgarno,
Phys. Rev. A {\bf 55}, R3299 (1997).

\bibitem{Verhaar} E. Tiesinga, B.~J. Verhaar, and H.~T.~C. Stoof, Phys. Rev.
A {\bf 41}, 4114 (1993).

\bibitem{vecqf1} Weiping Zhang, Phys. Lett. A {\bf 176}, 225 (1993);
Weiping Zhang, D.~F. Walls, and Barry Sanders, 
Phys. Rev. Lett. {\bf 72}, 60 (1994).

\bibitem{vecqf2} G. Lenz, P. Meystre and E.~M. Wright, 
Phys. Rev. Lett. {\bf 71}, 3271 (1993).

\bibitem{vecqf3} J. Javanainen, Phys. Rev. Lett. {\bf 75}, 1927 (1995);
H. D. Politzer, Phys. Lett. A 209, 160 (1995).

\bibitem{darkground} M. Born and E. Wolf, {\it Principles of Optics} 6ed.,
(Pergamon Press, 1980); M.~R. Andrews, M.-O. Mewes, N.~J. van Druten, D.~S.
Durfee, D.~M. Kurn, and W. Ketterle, Science {\bf 273}, 84 (1996).

\bibitem{photodetector} B. Saleh, {\it Photoelectron Statistics} 
	(Springer-Verlag, 1978).

\bibitem{size} I.~F. Silvera and J.~T.~M. Walraven, J. Appl. Phys. {\bf 52},
2304 (1981).

\end{thebibliography}
\end{document}